# Socializing the h-index

Graham Cormode[*]   Qiang Ma   S. Muthukrishnan   Brian Thompson

5/7/2013

**Abstract**

A variety of bibliometric measures have been proposed to quantify the impact of researchers and their work. The h-index is a notable and widely-used example which aims to improve over simple metrics such as raw counts of papers or citations. However, a limitation of this measure is that it considers authors in isolation and does not account for contributions through a collaborative team. To address this, we propose a natural variant that we dub the Social h-index. The idea is to redistribute the h-index score to reflect an individual's impact on the research community. In addition to describing this new measure, we provide examples, discuss its properties, and contrast with other measures.

**Keywords**: h-index, social impact,

# 1   Introduction

Across all academic disciplines, it is natural to want to measure the impact of an individual and his or her work. Consequently, many metrics have been proposed, based on properties of an individual's research output. These start with simple counts of papers (published in selective venues), or citation counts for papers, and become progressively more complex. These are used to compare the impact of individuals, influencing decisions around hiring and promotions. Given the attention such metrics receive, there has been much effort in designing them to be meaningful. For example, total paper counts give little indication of the quality of the work. Similarly, aggregate citation counts are distorted by a single highly-cited paper, and so do not indicate the subject's breadth.

Hirsch (2005) proposed the h-index: the largest integer $h$ such that the author has published at least $h$ papers with at least $h$ citations each. This measure has an intuitive appeal, and is not unduly influenced by a single high-impact paper, nor by a multitude of low-impact publications. Since then, a plethora of variations and alternative indices have been proposed to address perceived shortcomings of the h-index. Most of these measures evaluate an author solely based on his or her individual publication record. However, modern scientific research tends to be highly collaborative in nature. Consequently, we argue that new metrics are needed to reflect this reality. In this article, we introduce a measure which aims to capture the impact of a researcher not only on the research corpus, but also on his or her fellow researchers. Taking the h-index as a suitable metric for an individual's research impact, we can measure the contribution of a researcher on the community by the extent to which he or she boosts the h-indices of others. We formalize this notion with the definition of the Social h-index and demonstrate its properties. We perform a case study over the Computer Science research corpus, and show that it is distinct from other measures, and rewards more collaborative research styles.

# 2   Socializing the H-index

## 2.1   Definition and Properties

We define the Social h-index as a metric to reflect the impact of researchers on their community. We write $A(p)$ to denote the set of authors of paper $p$, and $P(a)$ to denote the set of papers authored by $a$. We use $h(a)$, the h-index of author $a$, and $H(a) \subseteq P(a)$, the set of papers that "support" the h-index of $a$, i.e. those that have at least $h(a)$ citations (this is similar to, but distinct from, the notion of the "h-core" of Rousseau (2006)). We first define a "contribution" function that counts the importance of a paper to the authors of that paper. That is, $\text{contrib}(p, a)$ measures the extent to which paper $p$ contributes to author $a$'s h-index. A natural instantiation is to set $\text{contrib}(p, a) = 1$ if $p$ supports $a$'s h-index, i.e. if $p \in H(a)$. However, due to ties, an author $a$ may have more than $h(a)$ papers with at least $h(a)$ citations. So for uniformity, we set $\text{contrib}(p, a) = h(a)/|H(a)|$ if $p \in H(a)$ and 0 otherwise.[†] This has the property that $\sum_{p \in P(a)} \text{contrib}(p, a) = h(a)$.

---

[*] Corresponding author, graham@dimacs.rutgers.edu

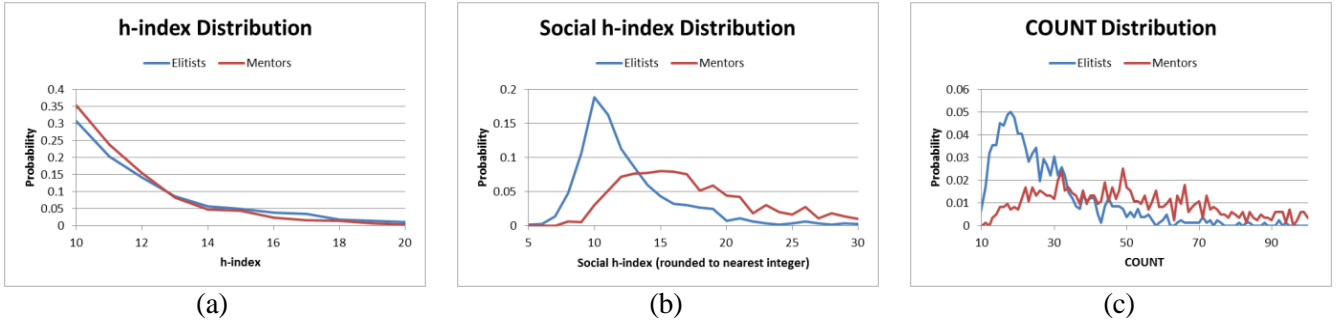

Figure 1: Distributions of (a) h-index, (b) Social h-index, and (c) COUNT for scientists who never work with 'novices' (elitists) and scientists for whom at least half of their collaborators are novices (mentors).

We define the Social h-index of an author $a$ to be the sum over all that author's papers of the (normalized) contributions to the paper's authors (including themselves). Then the Social h-index of $a$, $soc^h(a)$, is

$$soc^h(a) = \sum_{p \in P(a)} \frac{1}{|A(p)|} \sum_{a' \in A(p)} \text{contrib}(p, a').$$

Many variations of this definition are possible. We could choose the contrib function to give more credit for papers with higher citation counts; to evaluate the contribution of a paper based on the author's record at the time of publication; or to not reward an author for contributions to her own h-index. Based on our empirical study of these variations, they either gave broadly similar results, or had some undesirable properties, so we converge on this definition as the preferred instantiation of Social h-index. Note that, unlike h-index, Social h-index can decrease over time, when a paper which once contributed to one author's h-index ceases to do so. However, we observed that this rarely happens in real data, so the measure tends to increase over time.

With this choice of the contrib function, we have

$$\sum_a soc^h(a) = \sum_p \sum_{a \in A(p)} \frac{1}{|A(p)|} \sum_{a' \in A(p)} \text{contrib}(p, a') = \sum_p \sum_{a' \in A(p)} \text{contrib}(p, a') \sum_{a \in A(p)} \frac{1}{|A(p)|}$$

$$= \sum_p \sum_{a' \in A(p)} \text{contrib}(p, a') = \sum_a \sum_{p \in P(a)} \text{contrib}(p, a) = \sum_a h(a).$$

That is, the new measure preserves the sum of h-index values, but redistributes it among the authors. Observing this connection between $soc^h(a)$ and $h(a)$, we need to determine if there is a substantial difference between them. Is it possible that $soc^h(a) \approx h(a)$?

We show that this is not the case in practice by considering the behavior of researchers in Computer Science. First we collected data on authors and publications from a snapshot of the DBLP database on October 11, 2011. We then collected the citation history of each paper using Google Scholar to enable temporal analysis. This resulted in a dataset containing in total 1,017,553 authors and 2,764,012 papers.

We define a *novice* as an author of a paper with $h(a) = 0$ at the time of publication, and study different styles of collaboration with novices. From the data, we identify two groups of authors. The *elitists* are authors who have never published a paper with a novice, while the *mentors* are those for whom at least half of their coauthors are novices. Each group represents roughly 3.5% of the authors in DBLP.[‡] Figure 1(a) shows the distribution of h-index values across elitists and mentors. It is striking that the two distributions are extremely similar, indicating that h-index does not capture this aspect of collaboration style. In contrast, Figure 1(b) shows the distribution of Social h-index between the groups is quite different. The values for elitists peak much earlier, while mentors skew later. This reflects our intuition that Social h-index rewards those who encourage less experienced researchers.

---

[†] We observe that $H(a) = h(a)$ for 83% of all authors, and $H(a) \leq 1.25 \cdot h(a)$ for 95% of authors who have h-index $\geq 5$, so this choice of the contrib function does not significantly affect our analysis.

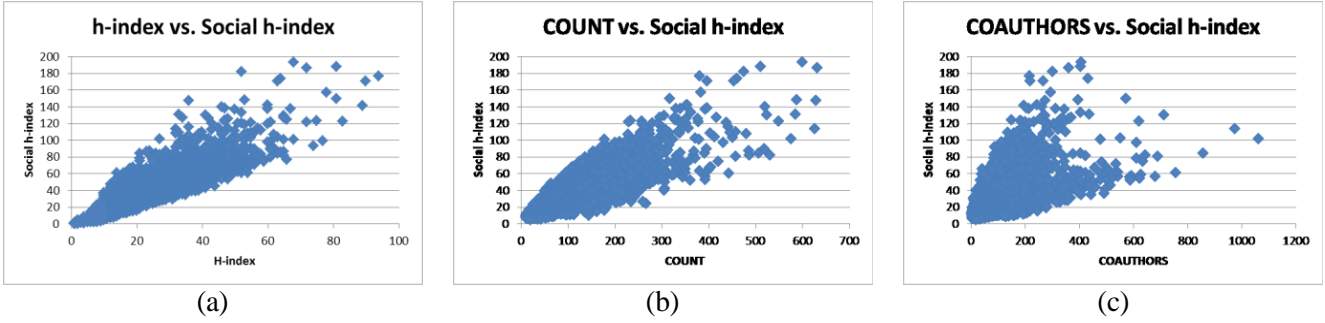

Figure 2: Comparing social h-index with other metrics: (a) h-index, (b) `COUNT`, and (c) `COAUTHORS`.

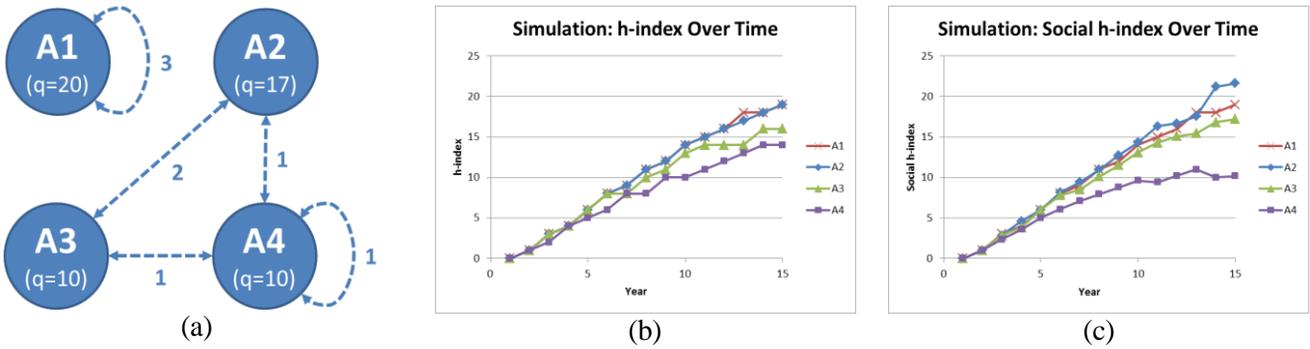

Figure 3: (a) Example collaboration graph $G$ used for the simulation; (b) h-index and (c) Social h-index of the four authors in $G$ over time.

It is also the case that the mentors tend to have more publications than their elitist colleagues, as seen in Figure 1(c) which plots the distribution of their publication `COUNT`. We also observe that they have more distinct coauthors (not shown). However, these measures do not usefully reflect their contribution to the field or their community.

We show correlation between Social h-index and these other measures for all authors in our dataset in Figure 2. While Social h-index is broadly correlated with h-index, number of papers (`COUNT`), and number of co-authors (`COAUTHORS`), there is much variation, indicating that this measure captures something quite distinct from these previously defined concepts.

## 2.2  Simulation

To better understand the relation between h-index and Social h-index, we perform a simple simulation exercise. We adopt the triangular peak-decay model of Guns and Rousseau (2009), where the annual rate of citations to a paper grows linearly to a peak, then decays linearly back to zero. In this simulation, each author $a$ has an inherent factor $q(a)$ that determines the number of citations his or her work receives: a paper $p$ will over time receive $\sqrt{\sum_{a \in A(p)} q(a)^2}$ citations (the $\ell^2$-norm of the $q$ values of the authors), reflecting the intuition that the quality of a paper is improved by each additional coauthor. Figure 3(a) shows an example collaboration graph $G$ consisting of four authors: each (weighted) edge indicates the number of papers per year produced by the linked pair of authors, with self-links indicating single-author papers. In this example, author A1 individually produces 3 papers a year, each of which receives 20 citations over time. A2 has the ability to individually produce papers that get 17 citations, but instead collaborates with other authors to jointly produce papers that over time receive 19 citations. As a result, the h-index of A2 closely follows that of A1, but never exceeds it (see Figure 3(b)). The Social h-index, on the other hand, additionally rewards the more "social" A2 for her role in furthering the careers of A3 and A4, who achieve a significantly higher h-index through the collaboration than they would if working individually (see Figure 3(c)). COUNT is not able to differentiate between any of the four authors, since they publish the same number of papers in this model.

---

[‡] To reduce bias from small samples, we only consider group members with h-index at least 10.

## 3  Data Analysis

| Rank($soc^h$) | Rank($h$) | Author | $soc^h$ | $h$ | Paper Count | Citation Sum | Distinct Coauthors |
|---|---|---|---|---|---|---|---|
| 1 | 13 | Thomas S. Huang | 193 | 68 | 600 | 20,102 | 406 |
| 2 | 5 | Jiawei Han | 188 | 81 | 511 | 27,663 | 405 |
| 3 | 11 | Philip S. Yu | 187 | 72 | 632 | 24,080 | 361 |
| 4 | 81 | Hans-Peter Seidel | 182 | 52 | 475 | 11,544 | 300 |
| 5 | 1 | Anil K. Jain | 177 | 94 | 380 | 34,501 | 216 |
| 6 | 23 | Alberto Sangiovanni-Vincentelli | 174 | 64 | 460 | 16,002 | 431 |
| 7 | 2 | Hector Garcia-Molina | 171 | 90 | 397 | 30,128 | 267 |
| 8 | 25 | Kang G. Shin | 171 | 63 | 453 | 14,505 | 217 |
| 9 | 7 | Donald F. Towsley | 157 | 78 | 383 | 20,264 | 294 |
| 10 | 5 | Ian T. Foster | 150 | 81 | 317 | 30,150 | 571 |
| 11 | 73 | Elisa Bertino | 148 | 53 | 588 | 12,153 | 395 |
| 12 | 517 | Chin-Chen Chang | 147 | 36 | 629 | 5,507 | 272 |
| 13 | 36 | Nicholas R. Jennings | 142 | 60 | 354 | 17,941 | 242 |
| 14 | 3 | Christos H. Papadimitriou | 142 | 89 | 351 | 27,761 | 195 |
| 15 | 160 | H. Vincent Poor | 140 | 46 | 520 | 8,392 | 262 |

Table 1: Top 15 authors in Social h-index

We apply Social h-index to citation data from Computer Science (obtained from Google Scholar and DBLP). Table 1 lists the 15 authors with highest Social h-index (rounded to the nearest integer), along with the corresponding h-index scores and other bibliometric statistics. We see that Social h-index rank is not determined by the other values shown. In particular, Hans-Peter Seidel is strictly dominated by Elisa Bertino in terms of h-index, paper count, citation sum, and number of distinct coauthors, yet has a higher Social h-index. This indicates that Social h-index considers subtleties in the publication record that are not reflected in the other metrics.

Nevertheless, we observe some trends among the top ranked authors. Many of them have a large total number of papers and many distinct coauthors, perhaps reflecting a history of fruitful collaborations and productive students. Discrepancies between h-index and Social h-index may reflect variations in subarea norms in collaboration styles (which are not explicitly adjusted for). For example, Christos Papadimitriou has the 3rd highest h-index, but falls to 14th place in the Social h-index ranking. This may result from his work in theoretical computer science, where researchers tend to work with fewer students than in other specialties.

Other researchers make substantial gains in the rankings when the Social h-index is applied. For instance, Chin-Chen Chang has the 12th highest Social h-index value in our dataset, but is ranked only 517th under h-index. His web page prominently lists that he has been the advisor of 49 Ph.D. students (29 graduated) and 106 masters students. Social h-index captures the fact that although he does not have as many very high-cited papers as some others, he has invested much of his time in helping young researchers get a successful start to their careers. More generally, all those who score highly on Social h-index have a large number of coauthors compared to the average number of 8. They have either mentored many graduate students, or had unusually many collaborations in industrial environments.

# 4 Conclusions

We have proposed the Social h-index as a way to measure the impact of a researcher on the academic community, taking into account the quantity and quality of publications, as well as the researcher's role in furthering the careers of other scientists through collaboration. We demonstrate that Social h-index is different from previously studied metrics, and can effectively distinguish between collaboration styles. We provide evidence of this through simulations as well as analysis of a large dataset of publications in the field of Computer Science.

There has been limited effort to capture such social effects before. Abbasi et al. (2010) proposed an index that rewards an author for collaborating with top researchers. This is in contrast to our approach, which rewards an author for helping less-prominent researchers as well. Kameshwaran et al. (2010) define a measure combining strength of publication record with eigenvector centrality to identify prominent researchers in the collaboration network, but the approach does not address their impact on others.

The notion of socialization of a metric can naturally be applied to other measures of academic success, such as the g-index (Egghe (2006)), and extended to count not only co-authors, but also the indirect influence on other researchers. Should such social measures become widely adopted and influential, it is natural that researchers will consciously or unconsciously start to "game" them. For example, Social h-index can be bolstered by adding junior researchers who have few publications as authors to a paper. It is then of interest to design measures which either prevent such manipulation, or which induce actions that genuinely benefit the community.